\documentclass[aps,prd,twocolumn,superscriptaddress]{revtex4-2}

\usepackage{dsfont}
\usepackage{physics}
\usepackage{amssymb}
\usepackage[makeroom]{cancel}
\usepackage{tikz}
\usepackage{hyperref}
\usepackage{xcolor}

\newcommand\myarray[1]{%
	\begingroup
	\renewcommand\arraystretch{1.33}
	\left\{ \begin{array}{@{}l@{}} #1 \end{array} \right.
	\endgroup}

\begin{document}


\title{Measuring correlations using local and nonlocal quenches}


\author{Alexey G. Mikhaylenko}
\email{mikhajlenko.ag@phystech.edu}
\affiliation{Lebedev Physical Institute, Russian Academy of Sciences, Moscow, Russia}
\affiliation{Skolkovo Institute of Science and Technology, Moscow, Russia}
\author{Andrew G. Semenov}
\affiliation{Lebedev Physical Institute, Russian Academy of Sciences, Moscow, Russia}
\affiliation{National Research University Higher School of Economics, Moscow, Russia}

\date{\today}

\begin{abstract}
We present a theoretical proposal for measuring correlation functions in a quantum field system based on the use of quantum quenches. Using the examples of massive and massless scalar field theories, we show that a short-term perturbation of the system leads to a dependence of the further evolution of the average field on the initial state's correlation functions. Considering two types of perturbations, namely local and nonlocal quenches, we show which correlation functions can be measured in each of these cases. The proposed procedure can be applied to study non-Gaussian correlations.
\end{abstract}


\maketitle


\section{Introduction \label{sec:intro}}
The state of a quantum system can be characterized by a set of correlation functions. Recently, progress in the experimental study of physical systems with a large number of degrees of freedom has made it possible to measure various correlation functions, including those of higher order \cite{chalopin2026connected,bureik2025suppression,schweigler2017experimental,rispoli2019quantum,armijo2010probing,endres2013single,xia2025gaussian,schweigler2021decay,murtadho2025systematic}. This progress is a step towards understanding the role of non-Gaussian correlations in the dynamics of quantum systems \cite{dolgirev2020non,van2025probing}.

In this paper, we propose a new procedure for measuring correlation functions in a quantum field system based on the use of quantum quenches.

The main idea is that a disturbance acting on a system for a short period of time can lead to a dependence of the evolution of observables on correlations of the initial state. Therefore, by measuring some observable after such an impact, it is possible to obtain information about correlation functions of various order. In this work we investigate the question of which correlation functions of the initial state become available for measurement after quenches of various types. Although the idea of measuring physical quantities by perturbation is not new \cite{cardy2011measuring,gluza2021recovering,froland2025entanglement,gritsev2007spectroscopy,menu2018quench,villa2019unraveling}, a specific feature of our work is that we provide a procedure by which a broad class of correlation functions can be measured. Therefore, this procedure provides another way to study non-Gaussian correlations.

Previously, the dynamical effects of interactions local in space, but not in time, in various spin \cite{bertini2016determination,reimann2026absence,torres2014local,torres2015relaxation,ganahl2012observation,dabelow2022thermalization,ashida2018solving,santos2020speck,eisler2008entanglement}, fermionic \cite{rossi2021signature,capizzi2023entanglement,bertini2017approximate,gouraud2023stationary,gouraud2022quench,ljubotina2019non,cavaliere2019coherent,rossi2022real}, and bosonic \cite{bastianello2019lack,del2022transport} systems were investigated. However, in this paper we explore the case of a local quench \cite{radovskaya2023local,PismaZhETF.124.152,PhysRevD.110.065011,PhysRevB.111.054312}, when the disturbance is localized not only in space, but also in time. Such perturbations were first investigated in the context of conformal field theory \cite{Calabrese_2007,Stephan_2011}, and then implemented as a flipping of several spins in spin systems \cite{fagotti2022global,karevski2002scaling,liu2014quench,fukuhara2013microscopic,jurcevic2014quasiparticle,yoshinaga2021ballistic} or as a quantum measurement \cite{bayat2018measurement}. The perturbation localized in space and time has already been realized experimentally in a quantum field simulator \cite{tajik2023experimental}. In addition, we propose a new type of quench, namely the nonlocal quench, and show which correlation functions it enables us to measure.

Although the proposed procedure is applicable to the study of non-Gaussian correlations of the initial state, in this paper, as a specific example, we consider the Gaussian initial state, namely, the ground state of a massive non-interacting theory. In this case all correlation functions can be calculated exactly. This enables us to obtain exact analytical results for the evolution of the average field after a local or nonlocal quench.

The outline of the paper is as follows. In Section \ref{sec:Main} we define a quantum quench and formulate the general idea. In Section \ref{sec:Local} we consider a local quench in massive and massless scalar field theory and describe the procedure for measuring correlations. In Section \ref{sec:Nonlocal} we generalize the procedure to a nonlocal quench. In Section \ref{sec:Conclusion} we summarize the results.

\section{General idea \label{sec:Main}}
A simple model that can be used to demonstrate the procedure for measuring correlations is a free real scalar field theory in a $d$-dimensional space given by a Hamiltonian
\begin{equation}
	\hat{H} = \frac{1}{2}\int d^dx \left( \hat{\pi}^2 + (\nabla\hat{\varphi})^2 + M^2\hat{\varphi}^2 \right).
\end{equation}
Here, the operators $\hat{\varphi}(\vec{x})$ and $\hat{\pi}(\vec{x})$ satisfy the canonical commutation relations
\begin{equation}
	[\hat{\varphi}(\vec{x}),\hat{\pi}(\vec{y})] = i\hbar\delta(\vec{x}-\vec{y}),
\end{equation}
\begin{equation}
	[\hat{\varphi}(\vec{x}),\hat{\varphi}(\vec{y})] = [\hat{\pi}(\vec{x}),\hat{\pi}(\vec{y})] = 0.
\end{equation}
In this paper, both massive ($M\neq0$) and massless ($M=0$) cases will be considered in detail.

Now let's describe what we mean by a quantum quench in this paper. We will call a quantum quench a perturbation that acts for a very short period of time and, in the limiting case, is described by a Hamiltonian $\hat{H}_{int}(t) = \delta(t-t_Q)\hat{V}$ where $t_Q$ is a moment of the quench and $\hat{V}$ is a Hermitian operator that will be specified later. As shown in Appendix \ref{sec:App_A}, as a result of such a perturbation, the state of the system described by density matrix $\hat{\rho}$ instantly changes to
\begin{equation}
	\hat{\rho}_Q = \hat{Q}\hat{\rho}\hat{Q}^\dagger
	\label{rho_Q}
\end{equation}
where the so called quench operator
\begin{equation}
	\hat{Q} = e^{-\frac{i}{\hbar}\hat{V}}.
	\label{Q}
\end{equation}

After the quench, the system continues to evolve with a free Hamiltonian $\hat{H}$, so that the observables at time $t>t_Q$ can be expressed using the retarded Green's function
\begin{equation}
	G^R(t,\vec{x}) = -\theta(t)\int\frac{d^dp}{(2\pi)^d}\frac{\sin(\omega_p t)}{\omega_p} e^{+i\vec{p}\vec{x}},
	\label{G_R_0}
\end{equation}
where $\omega_p=\sqrt{M^2+p^2}$. This function satisfies the equation
\begin{equation}
	(\partial_t^2-\Delta_x+M^2)G^R(t,\vec{x}) = -\delta(t)\delta(\vec{x}).
\end{equation}
For example, the evolution of the average field after the quench is expressed as follows
\begin{align}
	\left\langle\hat{\varphi}(t,\vec{x}) \right\rangle_Q  = &-\int d^dy~\big(\partial_t G^R(t-t_Q,\vec{x}-\vec{y})\cdot \left\langle \hat{\varphi}(\vec{y}) \right\rangle_Q +&&\notag\\& + G^R(t-t_Q,\vec{x}-\vec{y})\cdot \left\langle \hat{\pi}(\vec{y}) \right\rangle_Q \big).
	\label{1_point_function_evolution_tr}
\end{align}
Here and further, the triangular brackets $\langle\ldots\rangle$ and $\langle\ldots\rangle_Q$ denote the averaging with the density matrices $\hat{\rho}$ and $\hat{\rho}_Q$. If no time is specified, it is assumed that $t=t_Q$.

Now we are ready to formulate the main idea of this work. As can be seen from the formula (\ref{1_point_function_evolution_tr}), the evolution of the average field after the quench depends on the values of $\langle\hat{\varphi}\rangle_Q$ and $\langle\hat{\pi}\rangle_Q$ immediately after the quench. These averages, in turn, can be expressed in terms of averages calculated in the state $\hat{\rho}$ before the quench. The main idea of this work is that perturbations of the system by various operators $\hat{V}$ lead to the dependence of the evolution of the average field on various correlations of the initial state $\hat{\rho}$. This gives a simple algorithm for calculating correlations in a quantum state: (1) to act on the system using a certain type of quench and (2) to measure the average field after the quench. Then, depending on the type of $\hat{V}$, the average field will depend on certain correlations of the initial state $\hat{\rho}$, including non-Gaussian ones.

In the following sections, we will study in more detail the question of which correlations of the initial state $\hat{\rho}$ can be measured with specific implementations of the perturbation $\hat{V}$.

\section{Local quench \label{sec:Local}}
In this section, we will consider the first example of a quench, namely a local quench, when the perturbation is localized in space near a point $\vec{x}_Q$ and has the form \cite{radovskaya2023local}
\begin{equation}
	\hat{V} = \lambda \hat{\varphi}^n_S\left(\vec{x}_Q\right).
	\label{Local_quench}
\end{equation}
The operator $\hat{\varphi}$ is smeared near the point $\vec{x}_Q$ using a smearing function $\eta(\vec{x}-\vec{x}_Q)$ that decreases rapidly outside the small neighborhood of the point $\vec{x}_Q$ and satisfies the normalization condition $\int d^dx~\eta\left( \vec{x} - \vec{x}_Q \right) = 1$:
\begin{equation}
	\hat{\varphi}_S\left( \vec{x}_Q \right) = \int d^dx~\hat{\varphi}(\vec{x})\cdot\eta\left( \vec{x} - \vec{x}_Q \right).
	\label{smeared_operator}
\end{equation}
The smearing of each field individually, rather than the entire product, is explained by the need to avoid divergences from averages at coinciding points. This point is discussed in more detail in Appendix \ref{sec:App_B}.

The evolution of the average field after the quench is described by the formula (\ref{1_point_function_evolution_tr}). Let's express the averages $\langle\hat{\varphi}\rangle_Q$ and $\langle\hat{\pi}\rangle_Q$ immediately after the quench in terms of the averages before the quench. Expanding the exponents in the formula (\ref{rho_Q}), we obtain the following expression for the average value of an operator $\hat{A}$ immediately after the quench
\begin{align}
	\tr\big\{\hat{\rho}_{Q}&\hat{A}\big\} = \tr\big\{\hat{\rho}\hat{A}\big\} +&&\notag\\& +\sum_{k=1}^\infty \frac{1}{k!}\left(-\frac{i}{\hbar}\right)^k \tr\big\{ \hat{\rho} \big[\ldots \big[ \hat{A},\underbrace{\hat{V}\big],\ldots\hat{V}\big]}_k \big\}.
	\label{rho_Q_nonperturbative}
\end{align}
Then the 1-point averages immediately after the local quench are
\begin{equation}
	\left\langle \hat{\varphi}(\vec{y}) \right\rangle_{Q} = \left\langle \hat{\varphi}(\vec{y}) \right\rangle,
	\label{varphi_pert_1}
\end{equation}
\begin{equation}
	\left\langle \hat{\pi}(\vec{y}) \right\rangle_{Q} = \left\langle \hat{\pi}(\vec{y}) \right\rangle - \eta(\vec{y}-\vec{x}_Q) \cdot n \lambda \cdot \left\langle \hat{\varphi}_S^{n-1}(\vec{x}_Q)\right\rangle.
	\label{pi_pert_1}
\end{equation}
Substituting the expressions (\ref{varphi_pert_1}) and (\ref{pi_pert_1}) into the formula (\ref{1_point_function_evolution_tr}), we obtain the following expression for the evolution of the average field after the local quench in terms of correlations before the quench
\begin{align}
	\left\langle\hat{\varphi}(t,\vec{x}) \right\rangle_{Q(n)} &= \left\langle \hat{\varphi}(t,\vec{x}) \right\rangle + &&\notag\\+n&\lambda \cdot G_S^R(t-t_Q,\vec{x}-\vec{x}_Q) \cdot \left\langle \hat{\varphi}_S^{n-1}(\vec{x}_Q)\right\rangle.
	\label{average_evol_local}
\end{align}
The notation $Q(n)$ indicates the power of the operator $\hat{\varphi}$ in the local quench (\ref{Local_quench}). The first term on the right hand side corresponds to the evolution of the average field in the absence of the quench. The smeared retarded Green's function is defined as follows
\begin{align}
	G_S^R(t-&t_Q,\vec{x}-\vec{x}_Q) = &&\notag\\&=\int d^dy~ G^R(t-t_Q,\vec{x}-\vec{y})\cdot\eta(\vec{y}-\vec{x}_Q).
	\label{G_R_smeared}
\end{align}

As can be seen from (\ref{average_evol_local}), the evolution of the average field is entirely determined by the smeared Green's function. For this reason, this function will be examined in detail in Section \ref{sec:Smeared}.

\subsection{Measurement of correlations using local quenches \label{sec:Measurement}}
Let's use the formula (\ref{average_evol_local}) to express the correlation functions before the quench. Here we will assume that $\vec{x}=\vec{x}_Q$, that is, the correlation function will be determined by measuring the average field at the point of quench.
\begin{equation}
	\left\langle \hat{\varphi}_S^{n-1}(\vec{x}_Q)\right\rangle = \frac{\left\langle\hat{\varphi}(t,\vec{x}_Q) \right\rangle_{Q(n)} - \left\langle \hat{\varphi}(t,\vec{x}_Q) \right\rangle}{n\lambda G^R_S(t-t_Q,\vec{0})}
    \label{procedure}
\end{equation}
This formula is the basis for the procedure of measurement of correlation functions. By acting with local quenches (\ref{Local_quench}) with $n=2,3,4,\ldots$ and measuring the following evolution of the average field, correlation functions $\langle\hat{\varphi}_S\rangle$, $\langle\hat{\varphi}^2_S\rangle$, $\langle\hat{\varphi}^3_S\rangle$, $\ldots$ in the state $\hat{\rho}$ before the quench can be determined. It is easy to find the corresponding cumulants from these correlation functions and thereby investigate the non-Gaussian correlations of the initial state.

Thus, local quenches enable us to consecutively find correlations of the initial state. However, using a local quench of the form (\ref{Local_quench}), only averages of the form $\langle\hat{\varphi}_S^\alpha(\vec{x}_Q)\rangle$, $\alpha=1,2,3,\ldots$, can be measured, that is, the averages of the products of fields localized near the same point $\vec{x}_Q$. To get information about averages of a more general kind, we need to use another kind of quench, namely the nonlocal quench, which will be considered in Section \ref{sec:Nonlocal}.

Note that the idea to use perturbations local in space and time to measure physical quantities is not new. A recently developed method called local quench spectroscopy uses these kinds of perturbations to find the spectra of interacting systems \cite{villa2020local,millar2026quench}. Despite some similarities, including the need to measure only a local quantity (in our case, the average field), the procedure proposed in this paper differs from local quench spectroscopy, since it is aimed at obtaining information about correlations in the initial state, rather than about the dynamic properties of the system. In addition, in this paper, the choice of an observable for measurement does not depend on the specific type of local quench.

We would like to emphasize that the method proposed here is applicable in its current form only to non‑interacting theories. The reason is that in the presence of interaction, formula (\ref{1_point_function_evolution_tr}) is not valid, even if the interacting retarded Green’s function is used instead of $G^R$. Instead, the evolution of the average field in this case depends not only on $\langle\hat{\varphi}\rangle_Q$ and $\langle\hat{\pi}\rangle_Q$, but also on higher‑order averages. Developing a similar method for interacting theories is a separate important problem that requires the use of more advanced methods of nonequilibrium quantum field theory with non-Gaussian initial correlations \cite{mikhaylenko2026keldysh}.

\subsection{Smeared retarded Green's function \label{sec:Smeared}}
As noted above, the evolution of the average field is entirely determined by the smeared  retarded Green’s function. The goal of this section is to investigate the behavior of this function over a long period of time. For definiteness, we will use a Gaussian smearing function
\begin{equation}
	\eta(\vec{x}-\vec{x}_Q) = (2\pi\epsilon^2)^{-d/2} \exp\left\{ -\frac{(\vec{x}-\vec{x}_Q)^2}{2\epsilon^2} \right\}
	\label{gauss_smeared}
\end{equation}
with $\epsilon$ characterizing the size of the area in which the quench is localized. Substituting (\ref{G_R_0}) and (\ref{gauss_smeared}) into (\ref{G_R_smeared}) and taking the spatial integral, we get
\begin{align}
	&G^R_{S}(t-t_Q,\vec{x}-\vec{x}_Q) = -\theta(t-t_Q)\cdot&&\notag\\&\cdot\int\frac{d^dp}{(2\pi)^d}\frac{\sin(\omega_p(t-t_Q))}{\omega_p} e^{-\epsilon^2p^2/2+i\vec{p}(\vec{x}-\vec{x}_Q)}.
	\label{G_R_S_expl}
\end{align}
For convenience, we will assume that the measurement of the average field takes place at the same point where the quench was localized, that is, $\vec{x}=\vec{x}_Q$. Then the integral in the formula (\ref{G_R_S_expl}) can be reduced to an integral with respect to $p=|\vec{p}|$
\begin{align}
	&G^R_{S}(t-t_Q,\vec{0}) = -\theta(t-t_Q)\cdot&&\notag\\&\cdot\frac{S_{d-1}}{(2\pi)^d} \int_0^\infty dp~p^{d-1}\frac{\sin(\omega_p(t-t_Q))}{\omega_p} e^{-\epsilon^2p^2/2}.
	\label{G_R_S_expl_0}
\end{align}
Here $S_{d-1} = 2\pi^{d/2}/\Gamma(d/2)$~--- the area of a $(d-1)$-dimensional sphere in $d$-dimensional space, expressed in terms of the gamma function.

Let's find the asymptotic behavior of the smeared retarded Green's function at long times. We will consider the massive and massless cases separately.

\subsubsection{Massive case}
For $M\neq0$, the stationary phase method can be used to calculate the integral in the formula (\ref{G_R_S_expl_0}) at long times $t-t_Q\gg M^{-1}$ and $t-t_Q\gg \epsilon$. As a result, we get
\begin{align}
	&G^R_{S}(t-t_Q,\vec{0}) \approx &&\notag\\&-\frac{M^{d-1}}{(2\pi M(t-t_Q))^{d/2}}\cdot\sin\left(M(t-t_Q)+\frac{\pi d}{4}\right).
	\label{G_R_S_M_neq_0_asympt}
\end{align}
Substituting this expression into formula (\ref{average_evol_local}), we obtain the following asymptotic behavior of the average field at the point of the local quench
\begin{align}
    &\left\langle \hat{\varphi}(t,\vec{x}_Q) \right\rangle_{Q(n)} \approx \left\langle\hat{\varphi}(t,\vec{x}_Q) \right\rangle -n\lambda\cdot \left\langle \hat{\varphi}_S^{n-1}(\vec{x}_Q) \right\rangle\cdot&&\notag\\&~~~~~~\cdot \frac{M^{d-1}}{(2\pi M(t-t_Q))^{d/2}}\cdot\sin\left(M(t-t_Q)+\frac{\pi d}{4}\right).
    \label{average_field_big_time}
\end{align}
\begin{figure}[] 
	\centering
	\includegraphics[width=0.5\textwidth]{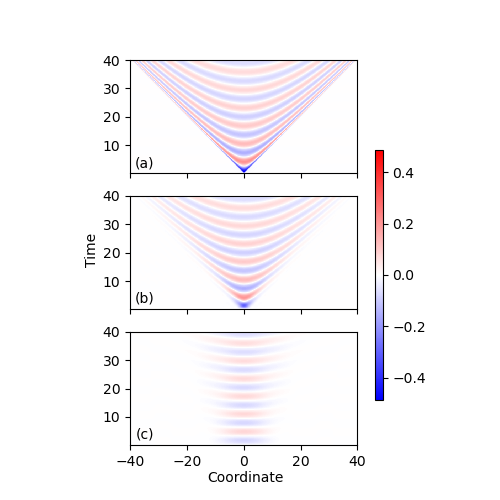}
	\caption{Dependence on the coordinate $M(x-x_Q)$ and time $M(t-t_Q)$ of the smeared retarded Green's function $G^R_S$ for a massive scalar theory in $d=1$ for the quench localization region $M\epsilon=0.2,1,5$(a,b,c).\label{G_S_M_neq_0_image}}
\end{figure}

The Figure \ref{G_S_M_neq_0_image} shows the evolution of the smeared rerarded Green's function (\ref{G_R_S_expl}) for a massive scalar theory in the dimension of space $d=1$ for three values of the size of the quench localization region $\epsilon$. It can be seen that over time, in all three cases, the behavior of $G^R_{S}(t-t_Q,x-x_Q)$ near the point $x_Q$ of the local quench becomes the same.

As was argued in \cite{radovskaya2023local}, the angle of the "light cone" is related to the maximum group velocity of the particles created during the quench
\begin{equation}
	v_{max} = \frac{\partial\omega_p}{\partial p}\bigg|_{p_{max}} = \frac{p_{max}}{\sqrt{M^2+p^2_{max}}} \backsim \frac{1}{\sqrt{1+(M\epsilon)^2}}.
\end{equation}
For a sufficiently small area of localization  $\epsilon\ll M^{-1}$ the disturbance propagates with the speed of light, whereas for $\epsilon\gg M^{-1}$ the front propagates more slowly.

\subsubsection{Massless case}
For $M=0$, the integral in the formula (\ref{G_R_S_expl_0}) can be expressed in terms of the (imaginary) error function
\begin{equation}
\text{erf}(x) = \frac{2}{\sqrt{\pi}}\int_0^x dy ~e^{-y^2},~\text{erfi}(x) = \frac{2}{\sqrt{\pi}}\int_0^x dy ~e^{+y^2}.
\label{erf}
\end{equation}
For physically relevant dimensions, the result is
\begin{align}
    &G^R_{S}(t-t_Q,\vec{0}) =-\theta(t-t_Q)\cdot&&\notag\\&\cdot\myarray{\frac{1}{2}\cdot\text{erf}\left(\frac{t-t_Q}{\sqrt{2}\epsilon}\right)~~(d=1)\\\frac{1}{2\sqrt{2\pi}\epsilon}\cdot\text{erfi}\left(\frac{t-t_Q}{\sqrt{2}\epsilon}\right)\cdot e^{-(t-t_Q)^2/2\epsilon^2}~~(d=2)\\\frac{1}{(2\pi)^{3/2}\epsilon^3}\cdot(t-t_Q)\cdot e^{-(t-t_Q)^2/2\epsilon^2}~~(d=3)}
\end{align}
In the big time limit $t-t_Q\gg\epsilon$
\begin{equation}
    G^R_{S}(t-t_Q,\vec{0}) \approx \myarray{-1/2~~(d=1)\\-1/2\pi(t-t_Q)~~(d=2)\\-\frac{t-t_Q}{(2\pi)^{3/2}\epsilon^3}\cdot e^{-(t-t_Q)^2/2\epsilon^2}~~(d=3)}
    \label{G_R_S_M_0_asympt}
\end{equation}
Thus in this limit the average field at the point of the local quench has the following asymptotic behavior (see formula (\ref{average_evol_local}))
\begin{align}
	&\left\langle \hat{\varphi}(t,\vec{x}_Q) \right\rangle_{Q(n)} \approx \left\langle\hat{\varphi}(t,\vec{x}_Q) \right\rangle -&&\notag\\&- n\lambda\cdot\left\langle \hat{\varphi}_S^{n-1}(\vec{x}_Q) \right\rangle\cdot \myarray{1/2~~(d=1)\\1/2\pi(t-t_Q)~~(d=2)\\\frac{t-t_Q}{(2\pi)^{3/2}\epsilon^3}\cdot e^{-(t-t_Q)^2/2\epsilon^2}~~(d=3)}
	\label{average_field_big_time_0}
\end{align}
\begin{figure}[] 
	\centering
	\includegraphics[width=0.5\textwidth]{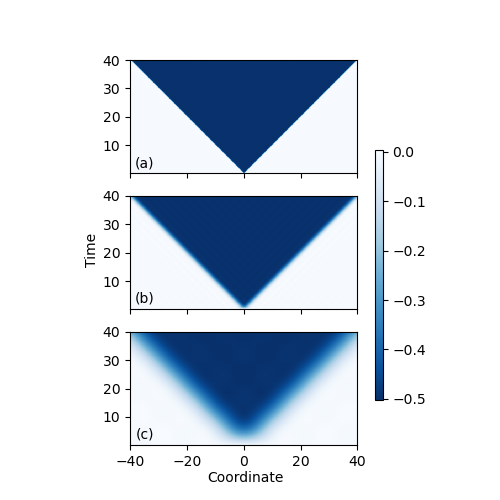}
	\caption{Dependence on the coordinate $\mu(x-x_Q)$ and time $\mu(t-t_Q)$ of the smeared retarded Green's function $G^R_S$ for a massless scalar theory in $d=1$ for the quench localization region $\mu\epsilon=0.2,1,5$(a,b,c). Here $\mu$ has dimension of mass and arbitrary positive value.\label{G_S_M_0_image}}
\end{figure}

Several remarks follow from this formula. First, in $d=1$, an important difference from the massive case is that the term that appears as a result of the quench does not become zero at $t\rightarrow\infty$, but takes a constant value throughout the entire space. Second, in $d=3$, the exponential decay of the smeared retarded Green’s function makes it challenging to use the method in the case of a massless scalar theory.

The Figure \ref{G_S_M_0_image} shows the evolution of the smeared rerarded Green's function (\ref{G_R_S_expl}) for a massless scalar theory in the dimension of space $d=1$ for three values of the size of the quench localization region $\epsilon$. It can be seen that, regardless of $\epsilon$, for $t-t_Q\gg\epsilon$, the function $G^R_{S}(t-t_Q,x-x_Q)$ takes a value $-1/2$ constant in space and time, consistent with the asymptotic expression (\ref{G_R_S_M_0_asympt}). 

Note that, in the massless case, the group velocity does not depend on the momentum and is always equal to the speed of light, so regardless of $\epsilon$, the light cone is always present.

\subsection{Local quench from the ground state \label{sec:Ground}}
In this section we will apply the results obtained above to study a specific example. As noted earlier, formula (\ref{procedure}) enables us to investigate non-Gaussian correlations of the initial state. However, in this section we will consider a simple Gaussian initial state. This will allow us to obtain an exact analytical solution for the evolution of the average field after the quench.

Let's assume that before the quench the system was in the ground state of a scalar theory with mass $M$, and at time $t=t_Q$ a local quench $\hat{V}=\lambda\hat{\varphi}_S^3(\vec{x}_Q)$ was applied. In the absence of the quench the average field is zero so the formula (\ref{average_evol_local}) has a simple form:
\begin{equation}
	\left\langle \hat{\varphi}(t,\vec{x}) \right\rangle_{Q(3)} = 3\lambda \cdot G_{S}^R(t-t_Q,\vec{x}-\vec{x}_Q) \cdot \left\langle \hat{\varphi}_S^2(\vec{x}_Q) \right\rangle.
\end{equation}

Since the full dependence on time and coordinate is contained only in the smeared retarded Green's function $G_{S}^R(t-t_Q,\vec{x}-\vec{x}_Q)$, the qualitative behavior of the average field after the quench does not differ from the behavior of the aforementioned function studied in the previous section and for $d=1$ shown in Figures \ref{G_S_M_neq_0_image} and \ref{G_S_M_0_image}. In $d=1$, this means that in the massive scalar theory at long times the average field oscillates with a decaying amplitude proportional to $1/\sqrt{t-t_Q}$. However, in the massless case the average field takes a constant value in space and time.

For sufficiently long time $t-t_Q\gg\epsilon$ and $t-t_Q\gg M^{-1}$ (for $M\neq0$) after the quench we can use the formulas (\ref{average_field_big_time}) and (\ref{average_field_big_time_0}) for the average field at the point of the quench
\begin{align}
    &\left\langle \hat{\varphi}(t,\vec{x}_Q) \right\rangle_{Q(3)} \approx -3\lambda\cdot\left\langle \hat{\varphi}^2_S(\vec{x}_Q) \right\rangle\cdot&&\notag\\&\cdot\myarray{\frac{M^{d-1}}{(2\pi M(t-t_Q))^{d/2}}\cdot\sin\left(M(t-t_Q)+\frac{\pi d}{4}\right)~~(M\neq0)\\1/2~~(M=0,~d=1)\\1/2\pi(t-t_Q)~~(M=0,~d=2)\\\frac{t-t_Q}{(2\pi)^{3/2}\epsilon^3}\cdot e^{-(t-t_Q)^2/2\epsilon^2}~~(M=0,~d=3)}
\end{align}

In both massive and massless cases, the magnitude of the average field is proportional to the smeared 2-point correlation function before the quench
\begin{align}
	\left\langle \hat{\varphi}^2_S(\vec{x}_Q) \right\rangle = &\int d^dx_1 \int d^dx_2~\left\langle \hat{\varphi}(\vec{x}_1) \hat{\varphi}(\vec{x}_2) \right\rangle \cdot&&\notag\\&\cdot \eta(\vec{x}_1-\vec{x}_Q) \cdot \eta(\vec{x}_2-\vec{x}_Q).
\end{align}
Correlation function for the ground state is
\begin{equation}
	\left\langle \hat{\varphi}(\vec{x}_1) \hat{\varphi}(\vec{x}_2) \right\rangle = \int \frac{d^dp}{(2\pi)^d} \frac{\hbar}{2\sqrt{M^2+p^2}} e^{+i\vec{p}(\vec{x}_1-\vec{x}_2)}.
	\label{2_point_func}
\end{equation}
In intermediate calculations, we will assume that the mass $M\neq0$ and will then take the limit $M\rightarrow0$ if needed. Taking into account the expression (\ref{gauss_smeared}) for the Gaussian smearing function, we obtain
\begin{align}
    &\left\langle \hat{\varphi}^2_S(\vec{x}_Q) \right\rangle = &&\notag\\&\myarray{\frac{\hbar}{4\pi}\cdot K_0\left(\frac{\epsilon^2M^2}{2}\right) \cdot e^{\epsilon^2M^2/2} ~~ (d=1) \\ \frac{\hbar}{8\sqrt{\pi}\epsilon} \cdot\left(1 - \text{erf}(\epsilon M)\right)\cdot e^{\epsilon^2M^2} ~~ (d=2) \\ \frac{\hbar M^2}{16\pi^2}\cdot\left( K_1\left(\frac{\epsilon^2M^2}{2}\right) - K_0\left(\frac{\epsilon^2M^2}{2}\right)\right)\cdot e^{\epsilon^2 M^2/2} ~~ (d=3)}
    \label{phi_S_2}
\end{align}
Here $K_\nu(x)$~ is a modified Bessel function of the second kind, $\text{erf}(x)$ is the error function given in (\ref{erf}). There are two limiting cases of the ratio between the size $\epsilon$ of the region in which the quench is localized and the characteristic scale $1/M$ in the system before the quench. In the limit $\epsilon M\ll1$:
\begin{equation}
    \left\langle \hat{\varphi}^2_S(\vec{x}_Q) \right\rangle \approx \myarray{-\frac{\hbar}{2\pi}\cdot\ln\left(\frac{\epsilon M}{2}\right)~~(d=1)\\\frac{\hbar}{8\sqrt{\pi}\epsilon}~~(d=2)\\\frac{\hbar}{8\pi^2\epsilon^2}~~(d=3)}
\end{equation}
These formulas are exact for $M=0$ and $d=2,3$. However, in $d=1$, an infrared regularization is needed for a massless scalar field. In the opposite limit $\epsilon M\gg1$:
\begin{equation}
    \left\langle \hat{\varphi}^2_S(\vec{x}_Q) \right\rangle \approx \frac{\hbar}{2M}\cdot\frac{1}{(2\sqrt{\pi}\epsilon)^d}.
\end{equation}

Similarly, the case of a quench from a thermal state can be considered. In this case the evolution of the average field will depend on the temperature $\beta^{-1}$.

For local quenches $\hat{V}=\lambda\hat{\varphi}_S^n(\vec{x}_Q)$ with $n>3$,  the evolution of the average field depends on the higher correlation functions of the initial state. For the ground state considered in this section, these correlation functions can be expressed in terms of 2-point correlation functions (\ref{phi_S_2}) using Wick's theorem. Therefore, the use of such quenches in this case does not bring new information. However, in the general case of a non-Gaussian initial state, such quenches may be of interest, since they make it possible to study non-Gaussian correlations.

\section{Nonlocal quench \label{sec:Nonlocal}}
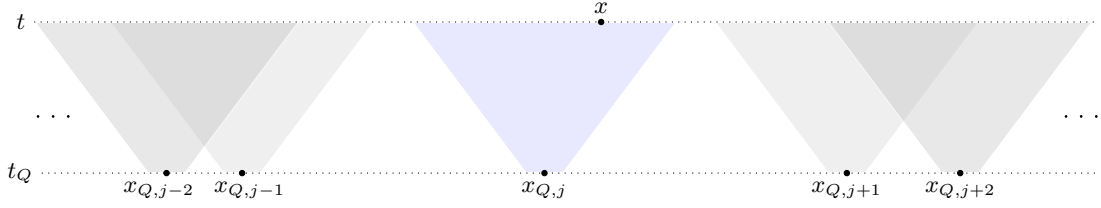
\begin{figure*}[] 
	\centering
	\begin{tikzpicture}
		\fill [fill=blue!30,draw=white,thick,opacity=0.3] (-1.25,0) -- (-0.75,0) -- (+0.75,2) -- (-2.75,2) -- (-1.25,0);
		\filldraw (-1,0) circle (1pt);
		\draw (-1,0) node[below]{$x_{Q,j}$};
		\fill [fill=black!20,draw=white,thick,opacity=0.3] (-5.25,0) -- (-4.75,0) -- (-3.25,2) -- (-6.75,2) -- (-5.25,0);
		\filldraw (-5,0) circle (1pt);
		\draw (-4.9,0) node[below]{$x_{Q,j-1}$};
		\fill [fill=black!30,draw=white,thick,opacity=0.3] (-6.25,0) -- (-5.75,0) -- (-4.25,2) -- (-7.75,2) -- (-6.25,0);
		\filldraw (-6,0) circle (1pt);
		\draw (-6.1,0) node[below]{$x_{Q,j-2}$};
		\fill [fill=black!20,draw=white,thick,opacity=0.3] (+3.25,0) -- (+2.75,0) -- (+1.25,2) -- (+4.75,2) -- (+3.25,0);
		\filldraw (+3,0) circle (1pt);
		\draw (+3,0) node[below]{$x_{Q,j+1}$};
		\fill [fill=black!30,draw=white,thick,opacity=0.3] (+4.75,0) -- (+4.25,0) -- (+2.75,2) -- (+6.25,2) -- (+4.75,0);
		\filldraw (+4.5,0) circle (1pt);
		\draw (+4.5,0) node[below]{$x_{Q,j+2}$};
		\draw[dotted] (-7.65,0) -- (6.3,0);
		\draw (-7.65,0) node[left]{$t_Q$};
		\draw[dotted] (-7.75,2) -- (6.3,2);
		\draw (-7.75,2) node[left]{$t$};
		\filldraw (-0.25,2) circle (1pt);
		\draw (-0.25,2) node[above]{$x$};
		\filldraw (-7.5,0.75) circle (0.3pt);
		\filldraw (-7.7,0.75) circle (0.3pt);
		\filldraw (-7.3,0.75) circle (0.3pt);
		\filldraw (5.9,0.75) circle (0.3pt);
		\filldraw (6.1,0.75) circle (0.3pt);
		\filldraw (6.3,0.75) circle (0.3pt);
	\end{tikzpicture}
	\caption{The case when the light cone with the origin at the point $\vec{x}_{Q,j}$ has not yet intersected with the light cone with the origin at any other point of nonlocal quench.\label{non_intersecting}}
\end{figure*}
As noted at the end of Section \ref{sec:Measurement}, local quenches only provide information about averages of the form $\langle \hat{\varphi}_S^\alpha(\vec{x}_Q) \rangle$ where all fields are localized near the same point $\vec{x}_Q$. To generalize the procedure to the measurement of more complicated correlation functions, we propose a new type of quench, namely a nonlocal quench 
\begin{equation}
	\hat{V} = \lambda \hat{\varphi}_S(\vec{x}_{Q,1})\ldots\hat{\varphi}_S(\vec{x}_{Q,n})
	\label{Nonlocal_quench}
\end{equation}
where each operator entering the perturbation $\hat{V}$ can be localized near its own point.

Using formulas (\ref{1_point_function_evolution_tr}), (\ref{rho_Q_nonperturbative}), and (\ref{Nonlocal_quench}), we obtain the following expression for the evolution of the average field after the nonlocal quench
\begin{align}
	\left\langle \hat{\varphi}(t,\vec{x}) \right\rangle_{Q(n)} =& \left\langle \hat{\varphi}(t,\vec{x}) \right\rangle + \lambda \sum_{j=1}^n G^R_{S}(t-t_Q,\vec{x}-\vec{x}_{Q,j})\cdot &&\notag\\\cdot\big\langle \hat{\varphi}_S&(\vec{x}_{Q,1}) \ldots\cancel{\hat{\varphi}_S(\vec{x}_{Q,j})}\ldots \hat{\varphi}_S(\vec{x}_{Q,n}) \big\rangle.
	\label{after_quench_non_local}
\end{align}

As can be seen from (\ref{after_quench_non_local}), in general, the evolution of the average field depends on $n$ correlation functions. In order to identify the contribution of only one of these correlation functions, it is necessary to measure the average field in the vicinity of the corresponding point after a sufficiently short time after the quench.

To better clarify this point, consider a situation where the time $t-t_Q$ after the quench is such that the light cone with the origin at the point $\vec{x}_{Q,j}$ has not yet intersected with the light cone with the origin at any other point near which the quench was made (see Figure \ref{non_intersecting}). This means that if $\vec{x}$ lies inside this light cone, then only one term remains in the formula (\ref{after_quench_non_local})
\begin{align}
	\left\langle \hat{\varphi}(t,\vec{x}) \right\rangle_{Q(n)} = &\left\langle \hat{\varphi}(t,\vec{x}) \right\rangle  + \lambda\cdot G^R_{S}(t-t_Q,\vec{x}-\vec{x}_{Q,j})\cdot&&\notag\\\cdot \big\langle \hat{\varphi}_S&(\vec{x}_{Q,1}) \ldots\cancel{\hat{\varphi}_S(\vec{x}_{Q,j})}\ldots \hat{\varphi}_S(\vec{x}_{Q,n}) \big\rangle.
	\label{non_local_1}
\end{align}

Formula (\ref{non_local_1}) is the basis for the procedure for measuring correlations using nonlocal quenches. As in the case of a local quench, measuring the average field after nonlocal quenches enables us to find correlations of the initial state. At the same time, due to the nonlocality of the interaction, it is possible to find the averages of the product of fields localized near different points.

Note that to use this method, it is sufficient to know the asymptotic behavior of the retarded Green’s function, which can be found by some other method, e.g., from the linear response. The sole requirement is that the Green’s function reaches it's asymptotic behavior before the time at which the light cones intersect.

For nonlocal quenches (\ref{Nonlocal_quench}) with $n>3$, the evolution of the average field depends on the higher-order correlation functions. This, in principle, makes it possible to study non-Gaussian correlations of the initial state.

Before moving on to the example, let us provide an interpretation of the nonlocal quench (\ref{Nonlocal_quench}). Note that formula (\ref{after_quench_non_local}) for the evolution of the average field contains only a term linear in $\lambda$. This means that the same expression can be obtained by expanding the exponential in the quench operator $\hat{Q}$ in a series up to the linear term:
\begin{equation}
    \hat{Q}=e^{-\frac{i}{\hbar}\hat{V}}\approx\hat{\mathds{1}} - \frac{i\lambda}{\hbar}\hat{\varphi}_S(\vec{x}_{Q,1})\ldots\hat{\varphi}_S(\vec{x}_{Q,n}).
\end{equation}
If, prior to the quench, the system was in the vacuum state $\ket{\Omega}$, then the state immediately after the quench,
\begin{equation}
    \ket{\Psi}_Q = \hat{Q}\ket{\Omega} \approx \ket{\Omega} - \frac{i\lambda}{\hbar}\hat{\varphi}_S(\vec{x}_{Q,1})\ldots\hat{\varphi}_S(\vec{x}_{Q,n})\ket{\Omega},
\end{equation}
is a superposition of two states. The second state is the result of particle creation near $n$ different points in space. Therefore, from the perspective of subsequent evolution of the average field, a nonlocal quench can be viewed as the creation of a superposition of this kind. A similar description applies to other initial states.

\subsection{Nonlocal quench from the ground state}
As an example, as in the Section \ref{sec:Ground}, we will consider a quench from the ground state of a scalar theory with mass $M$. In this case, the initial state is Gaussian and all the information about the state is contained in 2-point correlation functions, so there is no need to consider nonlocal quenches with $n>3$. Let's assume that at time $t=t_Q$ a nonlocal quench $\hat{V} = \lambda \hat{\varphi}_S(\vec{x}_{Q,1})\hat{\varphi}_S(\vec{x}_{Q,2})\hat{\varphi}_S(\vec{x}_{Q,3})$ was applied. Here the points $\vec{x}_{Q,1}$, $\vec{x}_{Q,2}$, $\vec{x}_{Q,3}$ are such that the localization regions do not intersect. In this case, the formula (\ref{after_quench_non_local}) gives
\begin{align}
	&\left\langle \hat{\varphi}(t,\vec{x}) \right\rangle_{Q(3)} = &&\notag\\&=\lambda \cdot G^R_{S}(t-t_Q, \vec{x}-\vec{x}_{Q,1}) \cdot \left\langle \hat{\varphi}_S(\vec{x}_{Q,2}) \hat{\varphi}_S(\vec{x}_{Q,3}) \right\rangle+ &&\notag\\&+\lambda \cdot G^R_{S}(t-t_Q, \vec{x}-\vec{x}_{Q,2}) \cdot \left\langle \hat{\varphi}_S(\vec{x}_{Q,1}) \hat{\varphi}_S(\vec{x}_{Q,3}) \right\rangle+ &&\notag\\&+ \lambda \cdot G^R_{S}(t-t_Q, \vec{x}-\vec{x}_{Q,3}) \cdot \left\langle \hat{\varphi}_S(\vec{x}_{Q,1}) \hat{\varphi}_S(\vec{x}_{Q,2}) \right\rangle.
	\label{after_quench_non_local_3}
\end{align}
Here we took into account that in the absence of a quench, the average field is zero, so the first term in the formula (\ref{after_quench_non_local}) is missing.

If during time $t-t_Q$ the light cones with the origins at points $\vec{x}_{Q,2}$, $\vec{x}_{Q,3}$, did not intersect with the light cone with the origin at point $\vec{x}_{Q,1}$, then for $\vec{x}=\vec{x}_{Q,1}$ only the first term contributes to the right-hand side of the equation (\ref{after_quench_non_local_3})
\begin{align}
	\left\langle \hat{\varphi}(t,\vec{x}_{Q,1}) \right\rangle_{Q(3)} =& &&\notag\\ =\lambda \cdot G^R_{S}&(t-t_Q, \vec{0}) \cdot \left\langle \hat{\varphi}_S(\vec{x}_{Q,2}) \hat{\varphi}_S(\vec{x}_{Q,3}) \right\rangle.
\end{align}
In addition, if $t-t_Q\gg\epsilon$ is satisfied, then the asymptotic expressions (\ref{G_R_S_M_neq_0_asympt}) and (\ref{G_R_S_M_0_asympt}) for the smeared retarded Green's function can be used.

A smeared 2-point correlation function at non-matching points can be calculated using (\ref{gauss_smeared}) and (\ref{2_point_func})
\begin{align}
	&\left\langle \hat{\varphi}_S(\vec{x}_{Q,2}) \hat{\varphi}_S(\vec{x}_{Q,3}) \right\rangle =&&\notag\\& =\int \frac{d^dp}{(2\pi)^d} \frac{\hbar}{2\sqrt{M^2+p^2}} e^{-\epsilon^2 p^2} e^{+i\vec{p}(\vec{x}_{Q,2}-\vec{x}_{Q,3})}.
\end{align}
\begin{figure}[]
	\centering
	\includegraphics[width=0.5\textwidth]{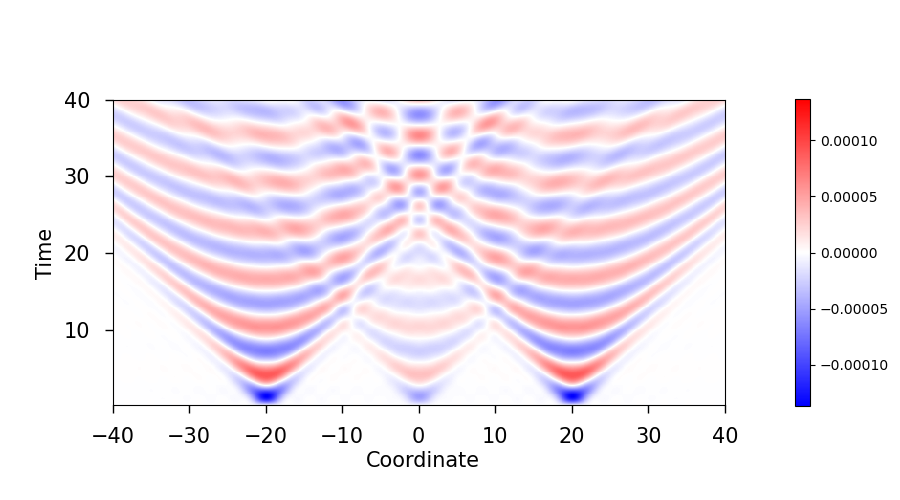}
	\caption{Dependence on the coordinate $M(x-x_Q)$ and time $M(t-t_Q)$ of the average field after the nonlocal quench for a massive scalar theory in $d=1$. Here $M\epsilon=1$, $\lambda=\hbar=1$.\label{G_S_non_locac_M_neq_0_image}}
\end{figure}
\begin{figure}[]
	\centering
	\includegraphics[width=0.5\textwidth]{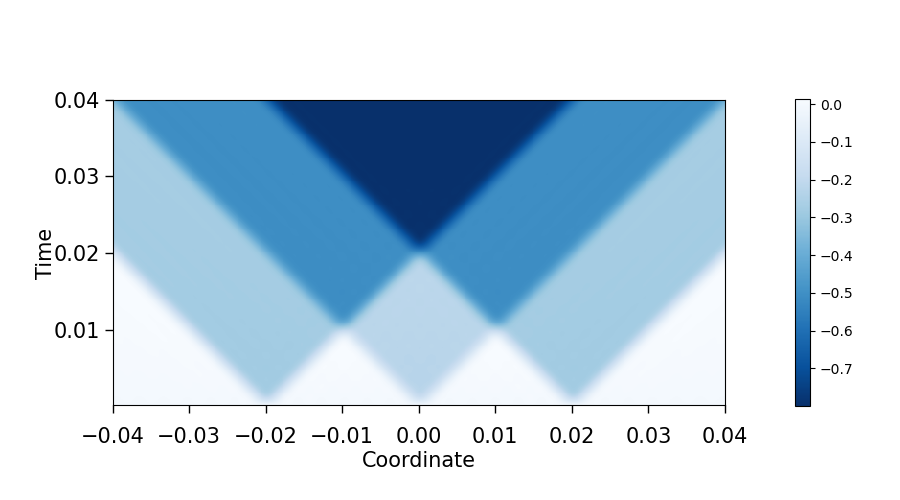}
	\caption{Dependence on the coordinate $\mu(x-x_Q)$ and time $\mu(t-t_Q)$ of the average field after the nonlocal quench for a massless scalar theory in $d=1$. Here $\mu$ is a momentum IR cutoff, $\mu\epsilon=10^{-3}$, $\lambda=\hbar=1$.}\label{G_S_non_locac_M_0_image}
\end{figure}

Figures \ref{G_S_non_locac_M_neq_0_image} and \ref{G_S_non_locac_M_0_image} show the evolution of the average field in $d=1$ after the nonlocal quench for the cases of massive and massless theory, respectively. The smaller absolute value of the average field in the light cone of the middle point is explained by the fact that the smeared 2-point correlation function decays with distance. This effect is more pronounced in the case of a massive theory, since in this case the correlation function decays exponentially with increasing distance, rather than logarithmically.

The use of the ground state allowed us to obtain an exact analytical solution for the evolution of the average field, and therefore served as a good example. However, we emphasize once again that the proposed procedure also enables us to analyze non-Gaussian initial states. In this case, the use of nonlocal quenches (\ref{Nonlocal_quench}) with $n>3$ makes it possible to study non-Gaussian correlations.

\section{Conclusion \label{sec:Conclusion}}
In this paper, a method for measuring correlations in a quantum field system using local and nonlocal quenches was proposed. The main idea was that a short-time perturbation on the system leads to a dependence of the further evolution of the observables on the correlations in the initial state. Taking the average field as an observable, we investigated the question of which correlation functions of the initial state can be measured using specific perturbations.

As the first example, the case of a local quench was considered. In this case the disturbance was localized in the vicinity of a certain point in space. The dependence of the evolution of the average field on correlations before the quench was found. It was shown how the consecutive application of local quenches makes it possible to find various correlation functions, including those of higher order. It has been demonstrated that a local quench enables us to obtain information only about correlations of a certain type, namely, the averages of the products of fields localized near the same point in space.

To overcome this problem, a new type of quench was proposed, namely, a nonlocal quench. In this case each operator entering the perturbation is localized near its own point in space and the averages of the product of fields localized near different points become available for measurement.

Both types of quenches have been analyzed for free massive and massless scalar field theories. The difference in the behavior of the average field over long periods of time for these two theories was investigated. In particular, it was shown that in the massless case in $1$-dimensional space, the average field gets a constant value in the entire space, which depends on the initial state of the system. A characteristic feature of the massive theory is that the size of the quench localization area affects the angle of the light cone, which is related to the maximum group velocity of the particles created during the quench.

In the future, we plan to apply the Keldysh diagram technique with non-Gaussian initial correlations \cite{mikhaylenko2026keldysh} to study the effect of interaction on the evolution of observables after quenches of various type. We expect that the presence of an interaction may erase information about the initial state due to long-term thermalization after a quench.

We hope that the proposed procedure for measuring correlation functions will contribute to the development of the field of the quantum state tomography and will be applied to study non-Gaussian correlations in many-body quantum systems.

\appendix
\section{Density matrix after quench \label{sec:App_A}}
To derive the formula (\ref{rho_Q}) describing the change in the density matrix as a result of quench, we assume that the interaction is activated for a very short but finite period of time from $t_Q -\Delta t/2$ to $t_Q+\Delta t/2$, and the Hamiltonian has the form
\begin{equation}
	\hat{H}(t) = \hat{H}_0 + f(t)\hat{V}.
\end{equation}
Here, the function $f(t)$ is non-zero only in the time interval mentioned above and satisfies the normalization condition
\begin{equation}
	\int_{t_Q-\Delta t/2}^{t_Q+\Delta t/2} dt~f(t) = 1.
	\label{f_condition}
\end{equation}
The evolution operator $\hat{U}$ in the presence of interaction satisfies the equation
\begin{equation}
	i\hbar \frac{\partial}{\partial t} \hat{U}\left( t, t_Q-\frac{\Delta t}{2} \right) = \hat{H}(t) \hat{U}\left( t, t_Q-\frac{\Delta t}{2} \right).
\end{equation}
Integrating this equation from $t_Q-\Delta t/2$ to $t$, iterating the resulting integral equation and putting $t = t_Q+\Delta t/2$ we get
\begin{align}
	&\hat{U}\left( t_Q+\frac{\Delta t}{2}, t_Q-\frac{\Delta t}{2} \right) = \hat{\mathds{1}} +&&\notag\\&+ \sum_{k=1}^{\infty} \left( -\frac{i}{\hbar} \right)^k \int_{t_Q-\Delta t/2}^{t_Q+\Delta t/2} dt_1 \int_{t_Q-\Delta t/2}^{t_1} dt_2 \ldots \int_{t_Q-\Delta t/2}^{t_{k-1}} dt_k~&&\notag\\&\hat{H}(t_1)\ldots\hat{H}(t_k).
\end{align}
Due to the condition (\ref{f_condition}) for $\Delta t\rightarrow0$, the contribution from $\hat{H}_0$ can be neglected
\begin{align}
	&\hat{U}\left( t_Q+\frac{\Delta t}{2}, t_Q-\frac{\Delta t}{2} \right) = &&\notag\\&=\hat{\mathds{1}} + \sum_{k=1}^{\infty} \left( -\frac{i}{\hbar} \right)^k \frac{1}{k!} \left( \int_{t_Q-\Delta t/2}^{t_Q+\Delta t/2} dt~f(t) \right)^k \hat{V}^k + o(1) = &&\notag\\&= e^{-\frac{i}{\hbar}\hat{V}} + o(1).
\end{align}
For $\Delta t\rightarrow0$, we get
\begin{equation}
	\hat{U}\left( t_Q+0, t_Q-0 \right) = e^{-\frac{i}{\hbar}\hat{V}} = \hat{Q}.
\end{equation}
This means that the instantaneous change in the density matrix as a result of quench is described by the formula (\ref{rho_Q}).

\section{Alternative definition of local quench \label{sec:App_B}}
To understand why (\ref{Local_quench}) was used for the definition of a local quench, let's consider an alternative definition
\begin{equation}
	\hat{V} = \lambda (\hat{\varphi}^n)_S(\vec{x}_Q) = \lambda \int d^dx~ \hat{\varphi}^n(\vec{x})\cdot\eta(\vec{x}-\vec{x}_Q).
\end{equation}
The 1-point correlation functions immediately after the quench can be found using the formula (\ref{rho_Q_nonperturbative})
\begin{equation}
	\left\langle \hat{\varphi}(\vec{y}) \right\rangle_{Q} = \left\langle \hat{\varphi}(\vec{y}) \right\rangle,
\end{equation}
\begin{equation}
	\left\langle \hat{\pi}(\vec{y}) \right\rangle_{Q} = \left\langle \hat{\pi}(\vec{y}) \right\rangle - \lambda n\cdot \left\langle \hat{\varphi}^{n-1}(\vec{y}) \right\rangle \cdot \eta(\vec{y}-\vec{x}_Q).
\end{equation}
Hence, the evolution of the average field after the quench is described by the following expression
\begin{align}
	\left\langle \hat{\varphi}(t,\vec{x}) \right\rangle_{Q(n)} =& \left\langle \hat{\varphi}(t,\vec{x}) \right\rangle + \lambda n \int d^dy~ G^R(t-t_Q,\vec{x}-\vec{y})\cdot &&\notag\\&\cdot\left\langle \hat{\varphi}^{n-1}(\vec{y}) \right\rangle\cdot\eta(\vec{y}-\vec{x}_Q).
\end{align}
This expression contains the average values of the product of fields at the coinciding points. Generally speaking, such averages diverge. Let's demonstrate this using the example of the thermal state in the absence of interaction, when the 2-point average has the following form
\begin{align}
	&\left\langle \hat{\varphi}(\vec{x}_1)\hat{\varphi}(\vec{x}_2) \right\rangle = &&\notag\\&=\int\frac{d^dp}{(2\pi)^d} \frac{\hbar}{2\omega_p}\coth\left(\frac{\beta\hbar\omega_p}{2}\right) e^{+i\vec{p}(\vec{x}_1-\vec{x}_2)},
	\label{phi_phi_thermal}
\end{align}
where $\omega_p=\sqrt{M^2+p^2}$. At coinciding points
\begin{equation}
	\left\langle \hat{\varphi}^2(\vec{x}) \right\rangle \propto \int_0^\infty dp~\frac{p^{d-1}}{\omega_p} \coth\left(\frac{\beta\hbar\omega_p}{2}\right).
\end{equation}
With $p\rightarrow\infty$, the integrand behaves like $p^{d-2}$, so there is an ultraviolet divergence in any dimension $d$ of the space. One way to eliminate this divergence is to use the normally-ordered perturbation operator $\hat{V}$\cite{belkovich2026global}.

On the other hand, this problem is not present when using the definition (\ref{Local_quench}) for the local quench \cite{radovskaya2023local}. Indeed, as can be seen from the formula (\ref{average_evol_local}), in this case, the expression for the evolution of the average field after quench contains the average values of the product of the smeared fields. The reason why such averages do not diverge is easy to understand using the same example of a 2-point average, but now with smeared fields
\begin{align}
	\left\langle \hat{\varphi}^2_S(\vec{x}_Q) \right\rangle = &\int d^dx_1 d^dx_2~\left\langle \varphi(\vec{x}_1)\varphi(\vec{x}_2) \right\rangle\cdot&&\notag\\&\cdot\eta(\vec{x}_1-\vec{x}_Q) \cdot \eta(\vec{x}_2-\vec{x}_Q).
\end{align}
After substituting the expression (\ref{phi_phi_thermal}), the Fourier transform from the smeared function appears in the integral expression. Let the smearing function be Gaussian (see formula (\ref{gauss_smeared})). Then its Fourier transform is also a Gaussian function, so the integrand decreases exponentially at $p\rightarrow\infty$, and there is no ultraviolet divergence.

\begin{acknowledgments}

This work was supported by the Russian Science Foundation under grant No. 25-22-00832.

\end{acknowledgments}

\bibliography{BiBTeX_File}

\end{document}